# A New Spin Gapless Semiconductors Family: Quaternary Heusler Compounds

G. Z. Xu,[1] E. K. Liu,[1] Y. Du,[1] G. J. Li,[1] G. D. Liu,[2] W. H. Wang,[1,a)] and G. H. Wu[1]

[1]*Beijing National Laboratory for Condensed Matter Physics, Institute of Physics, Chinese Academy of Sciences, Beijing 100190, P. R. China*

[2]*School of Material Science and Engineering, Hebei University of Technology, Tianjin 300130, P. R. China*

**Abstract:**

Using first-principles calculations, we investigate the band structures of a series of quaternary LiMgPdSn-type Heusler compounds. Our calculation results show that five compounds CoFeMnSi, CoFeCrAl, CoMnCrSi, CoFeVSi and FeMnCrSb possess unique electronic structures characterized by a half-metallic gap in one spin direction while a zero-width gap in the other spin direction showing spin gapless semiconducting behavior. We further analysis the electronic and magnetic properties of all quaternary Heusler alloys involved, and reveal a semi-empirical general rule (total valence electrons number being 26 or 28) for indentifying spin gapless semiconductors in Heusler compounds. The influences of lattice distortion and main-group element change have also been discussed.

PACS numbers: 71.20.-b, 71.20.Lp, 75.50.Cc

**Key words:** quaternary Heusler compounds, Spin gapless semiconductors, Half-metallic ferromagnets

Spin gapless semiconductors (SGS), as a new concept in spintronics, was first proposed theoretically and verified experimentally in doped Pd-based oxide material by Wang in 2008.[1,2] It can be regarded as a combination of gapless semiconductors[3] and half-metallic (HM) ferromagnets.[4] In gapless semiconductors, such as the first studied HgCdTe and HgCdSe, etc.,[3] or the recently widely studied graphene,[5] no threshold energy is required to excite the carriers from valence states to conduction states owing to the zero-width gap, thus achieving considerably higher electron mobility and more sensitive response to the external fields than the ordinary semiconductors. HM ferromagnets, in which conducting electrons are 100% spin polarized, are also attractive novel spintronic materials with an insulating or semiconducting gap in one spin direction, while in the other the electrons show a metallic behavior. In the case of SGS, for which density of states (DOS) scheme is shown in Fig.1 (a), the spin down gap across the Fermi level retains as in HM ferromagnets, but a zero-width gap appears in the spin up direction, where the electrons show a gapless semiconducting behavior. The special DOS feature of SGS indicates some novel transport properties and applications in spintronic devices, considering that the conducting electrons or holes are not only 100% spin polarized but also easily excited. The design of SGS can either by introducing magnetism in gapless semiconductors[1, 6, 7] or by opening a zero bandgap in one spin channel based on the ferromagnetic semiconductors[8] or HM ferromagnets.[9, 10]

Recently, Heusler alloy $Mn_2CoAl$, which was previously calculated to be a HM ferromanget,[11] was experimentally demonstrated to be a SGS material.[9] Later on,

Galanakis *et al.*[10] calculated a series of inverse Heusler compounds and found several of them to be candidates of SGS. Heusler compounds[12] are a huge family with more than 1000 members, having been applied in many areas owing to their multifunctionity and rich properties.[13] They have innate advantages in spintronic devices due to the good compatibility with conventional semiconductors. In the present paper, we propose a new material family that possesses the properties of SGS: quaternary Heusler compounds.

Quaternary Heusler can be designated by XX'YZ with 1:1:1:1 stoichiometry, where X, X', Y are transition metals and Z are main group elements. XX'YZ crystallizes in a LiMgPdSn or called *Y*-type structure[14] (as shown in Fig. 1 (b)) with space group of $F\bar{4}3m$，where the X and X' atoms with more valence electrons occupy the 4a (A) and 4b (C) sites, respectively, the Y atoms with less valence electrons occupy the 4c (B) sites, the main group element Z lies at the 4d (D) sites. The reliability of this occupation rule in quaternary Heusler alloys has been proved by both theoretical and experimental studies.[15, 16] Those studies also imply that the XX'YZ alloy tends to form highly ordered structure rather than the disorder configuration. Based on the above structure, we have carried out systematic first-principles calculations by changing X, X' from V to Co and Y from Ti to Mn along the Periodic Table, while choosing Z as Al or Si. Our calculations used the CASTEP package[17, 18] based on the pseudopotential method with a plane-wave basis set. The exchange correlation energy was treated under the generalized gradient approximation (GGA).[19] For all cases a plane-wave basis set cut-off energy of 400 eV and a mesh of $10\times10\times10$ k-points were employed to ensure good convergence. By performing the geometry optimization calculation, we obtained the equilibrium lattice constants.

As a result, our calculations have indentified four compounds that would be probable SGS: CoFeMnSi, CoFeCrAl, CoMnCrSi, CoFeVSi. The equilibrium lattice constants and atom-resolved magnetic moments are presented in Table S1. The nearly integral magnetic moments convince the existence of the HM gap, which is a prerequisite of being SGS. In addition, these alloys can be synthesized easily as they can be regarded as combination of full Heusler alloys $X_2YZ$ and $X'_2YZ$. For example, CoFeMnSi can be seen as combination of $Co_2MnSi$ and $Fe_2MnSi$, which has been successfully synthesized by Dai *et al.*[20] The DOS of all these four alloys exhibit a band gap in one spin channel and a typical energy valley approaching zero at the Fermi energy in the other spin channel, which fulfill the requirements of SGS as above described. Spin resolved band structure and DOS of CoFeCrAl is given in Fig.2 as representation (DOS of the other three are presented in Figure S2). An energy gap is opened in the minority-spin state across the Fermi energy as in a HM ferromagnet, the most eye-catching point is that in the majority-spin state the energy band touches the Fermi level at *K*, *L* (valence band) and *X* (conduction band) points in the Brillouin zone, which corresponds to a valley in the DOS at the Fermi energy. This closed bandgap character in the majority-spin state suggests that CoFeCrAl is a SGS rather than a normal HM ferromagnet, and further experimental studies are deserved to confirm it.

In order to illustrate more detailed the properties of quaternary Heusler compounds and search the possible rule of finding SGS that underneath the superficial results. In Fig. 3, we listed the calculated results in a two dimensional graph taking calculated equilibrium lattice parameters as one variable and total valence electrons the other. We know that if there exists a bandgap in one spin direction, the magnetic moment of the compound is supposed to be integer and obey the Slater-Pauling rule.[21,22] In our cases, the magnetic moments of the most compositions involved follow the generalized Slater-Pauling rule of $M=N_v-24$ or $M=N_v-18$,[23] indicted by the different background in Fig. 3. There are some of the compounds containing V or Ti as Y atom (marked by squares with a hollow) deviate from the Slater-Pauling rule with smaller magnetic moments comparing with the estimated integer value. Divided by the total valence electrons of 21, there are 12 or 9 spin-down electrons occupy below the Fermi energy, which suggests that systems with 24, 21 and 18 valence electrons are usually nonmagnetic, in consideration of the usual incompatibility of semiconducting and magnetism in one single phase alloy. Remarkably, we noticed that the four compounds that were supposed to be SGS all have valence electrons of 26 or 28, and the previous $Mn_2CoAl$ was also 26 electron system.[9] In order to justify it is not coincidental but rather an effective criterion for being SGS, in what follows we carried out detailed discussions from the perspective of atomic hybridization.

Previously, there have been elaborated analysis of atomic hybridization and the origin of HM band-gap in the studies of Heusler-based HM ferromagnets.[24] Here for the sake of searching for SGS, one should take into account orbital hybridizations of both the spin up and spin down directions. For comparison, in Fig. 4, we still present the schematic diagram of spin down hybridization. In the system obeying the Slater-Pauling rule of $M=N_v-24$, the Fermi level is supposed to locate between the $3t_{1u}$ and $2e_u$ degenerate levels as shown in the figure. The gap between $3t_{1u}$ and $2e_u$, i.e., energy gap of spin down electrons around the Fermi level, is basically determined by the $d$-orbital hybridizations between X atom at A site and X' atom at C site. In spin up direction, the total number of energy levels and symmetry representations are identical to spin down direction, however, the relative position of the hybridization energy scale is moved by the exchange splitting in atoms both inside and between. Combining the results of all the band structures that we have obtained, the hybridization picture are proved to be similar except that in some systems $2e_u$ and $3t_{1u}$ may intersect at the Γ point (not shown in the figure). According to the above-described hybridization scheme, we index the $d$-orbital hybridized bands of CoFeCrAl in Fig. 2 using the corresponding representations. Now let's re-examine the condition of being SGS, which also needs a gap exist around the Fermi level as in the spin down state, only that the bandgap width is zero. We conclude that the way to guarantee the Fermi level falls in between a band gap is that it should not locate in those degenerate bands, otherwise it must cut the band without a gap. In other word, it requires the total valence electrons to be 26 or 28, with the Fermi level locating in between $2e_u$ and $2e_g$ (shown in Fig. 4) or $2e_g$ and $3t_{2g}$ orbitals. We can take the band structure of CoFeCrAl (Fig. 2) as an example to interpret this. There, the Fermi level was located between the $2e_g$ and $2e_u$ orbitals, and in the meantime, the top of $2e_u$ bands and the bottom of $2e_g$ bands rightly touch at the Fermi level, thus showing SGS properties. If the total valence electrons were subtracted by one or added by one, the Fermi level would move down or up. It is because that the $2e_u$ or $2e_g$ is degenerated in some $k$ points, the Fermi level must cut the entangled bands in order to contain one more or less electron. In this sense, the compounds with 25 or 27 valence electrons cannot be the candidates of SGS. How about locating the Fermi energy between the lower degenerate orbitals, i.e., $2e_u$ and $3t_{1u}$ or $3t_{1u}$ and $3t_{2g}$? They can be excluded from our consideration because they correspond to the total valence electrons of 24 and 21, which are usually nonmagnetic according to the above discussion. Finally we can make a conclusion that the prerequisite of finding SGS in this quaternary Heusler system is rendering the total valence electron number to be 26 or 28. Based on this rule, we have calculated another compound with 26 valence electrons: FeMnCrSb, and as we expected, its DOS structure (shown in Figure S2) reveals the character of SGS.

Moreover, the above rule is not only reasonable in quaternary Heusler alloys what we concerned in this work, it can be also applied to other Heusler compounds as long as the unit cell containing four atoms; $Mn_2CoAl$ is such a case. Nevertheless, it should be noted that the valence electron number is a necessary but not sufficient condition to indentify SGS. For instance, $Fe_2MnAl$ has 26 valence electrons but it is not a half metal,[25] therefore the fundamental condition is not fulfilled; $Co_2MnAl$[26] is a well known HM ferromagnets with 28 electrons, but in spin up direction, the upper and lower band were both cut by the Fermi level. Finally, we should point out here that the unique band structure of quaternary Heusler SGS can be tuned by stretching or contracting or even tetragonally distorting the lattice. The band structure of CoMnCrSi with equilibrium lattice parameters and uniaxial distorted lattice are exhibited in the upper and lower panel of Fig. 5(a), respectively. Ignoring the Brillouin zone change caused by the broken cubic symmetry, it can be seen clearly that the spin up band rightly touches the Fermi level after distortion, though the spin down gap becomes a little smaller as well. On the other hand, the substitution of *sp*-elements by atoms in the same group can also destroy the electronic structure of SGS. For instance, as shown in Fig. 5b, when we use isoelectronic Ga or In to replace Al in CoFeCrAl, the spin up band mainly keeps unchanged while the spin down gap is destroyed due to the overlapping of $3t_{1u}$ and $2e_u$, which can be attributed to the weakening of the covalent hybridization between the main-group and transition atoms.

In conclusion, we have performed first-principles calculations on a series of quaternary Heusler compounds and finally found five compounds: CoFeMnSi, CoFeCrAl, CoMnCrSi, CoFeVSi and FeMnCrSb that are identified as potential candidates of SGS from the DOS and band structure character. The former four compounds ought to be easily synthesized and high Curie temperatures are expected when regarding them as a combination of $X_2YZ$ and $X'_2YZ$, making them to be promising spintronic devices. In addition, a general rule of valence electron number being 26 or 28 is summarized in search for SGS among Heusler compounds. This semi-empirical rule can narrow the scope of finding SGS in future. Nevertheless, as it is a necessary but not sufficient condition, specific systems should be treated differently and the alteration of *sp* atoms in the same group may cause considerable change of electronic properties.

**Acknowledgements**

This work was supported by National Natural Science Foundation of China (Grant Nos. 51171207 and 51021061) and National Basic Research Program of China (973 Programs: 2012CB619405).

**Figure captions:**

**FIG. 1** (Color online). (a) Density of states (DOS) scheme of half-metallic ferromagnets (left) and spin gapless semiconductors (right). Filled areas represent occupied states, and the arrows indicate majority ($\uparrow$) spin and minority ($\downarrow$) spin; (b) Crystal structure of *Y*–type quaternary Heusler compound, the atom X, X' occupies the Wyckoff position of 4a (0, 0, 0) and 4b (0.5, 0.5, 0.5), respectively, Y atom locates at 4c (0.25, 0.25, 0.25) and Z at 4d (0.75, 0.75, 0.75).

**FIG. 2** (Color online). Part of band structure and density of states (DOS) of CoFeCrAl: (a) majority spin, the green lines are used for emphasize, (b) density of states, (c) minority spin. The irreducible representations of *d*-orbital hybridized bands are given for the $\Gamma$ point, which are in accordance with those provided in Fig.4.

**FIG. 3** (Color online). A presentation of all quaternary Heusler compounds that calculated in this work, with theoretical equilibrium lattice parameters as one axis and total valence electrons $N_v$ as the other. The different background stands for the systems obeying the Slater-Pauling rule of $M=N_v-18$ (light yellow) and $M=N_v-24$ (light blue), respectively. The compounds with 21 electrons lie in the boundary are usually nonmagnetic, belonging to neither region. The squares with hollow represent the systems that deviate from the rule with non-integer magnetic moments. The red stars mark the compositions that are suggested to be SGS, and $Mn_2CoAl$ is shown for comparison.

**FIG. 4** (Color online). The sketch of possible hybridizations between *d* orbital of transition atoms at different site in the XX'YZ quaternary Heusler compound (left for spin-down electrons, right for spin-up electrons). The symmetry representations of degenerated orbitals refer to the work by Galanakis *et al.*[21] The green line indicates the location of the Fermi level, which is actually in the same position to both spin directions.

**FIG. 5** (Color online). Band structures of CoMnCrSi and CoFeCrGa/In, the blue (red) line represents the spin up (down) electron. (a) The upper is the band structure of CoMnCrSi with equilibrium lattice constants, the lower one is with a uniaxial expansion of 4% while keeping the volume unchanged (the *k* path is different from the cubic one due to the reduced symmetry). (b) Partial band structures of CoFeCrGa and CoFeCrIn under equilibrium lattice constants.

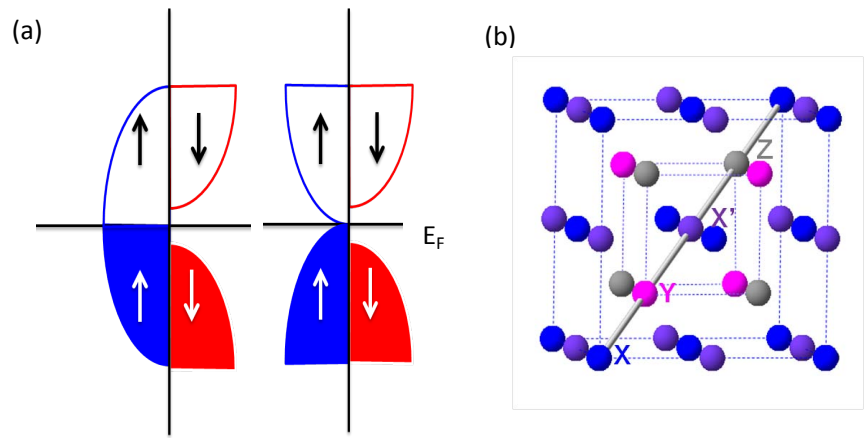

Figure 1

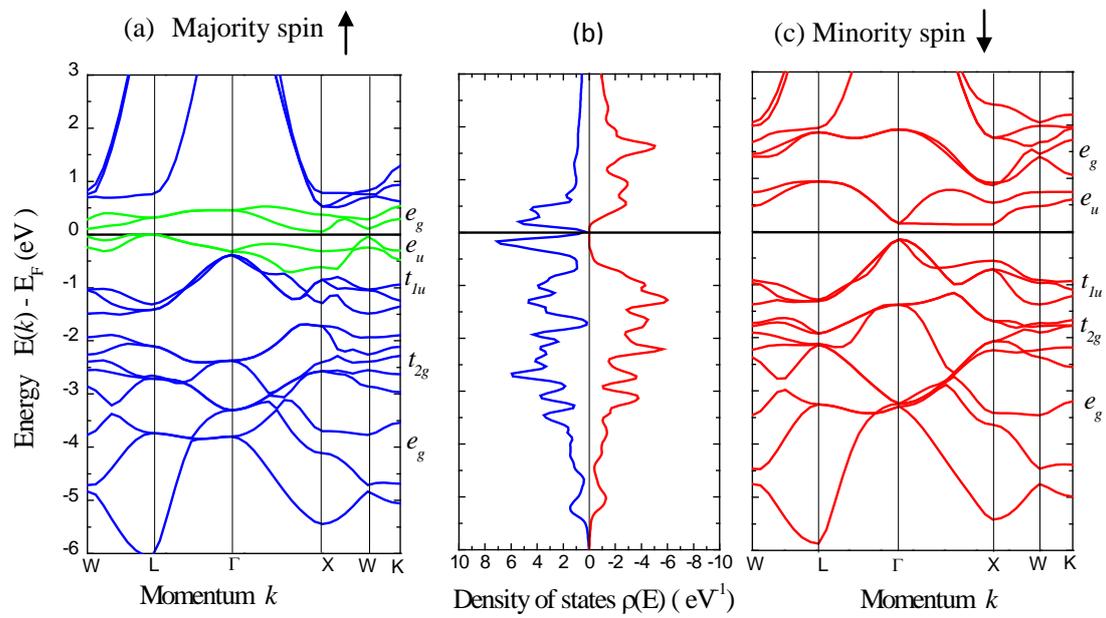

Figure 2

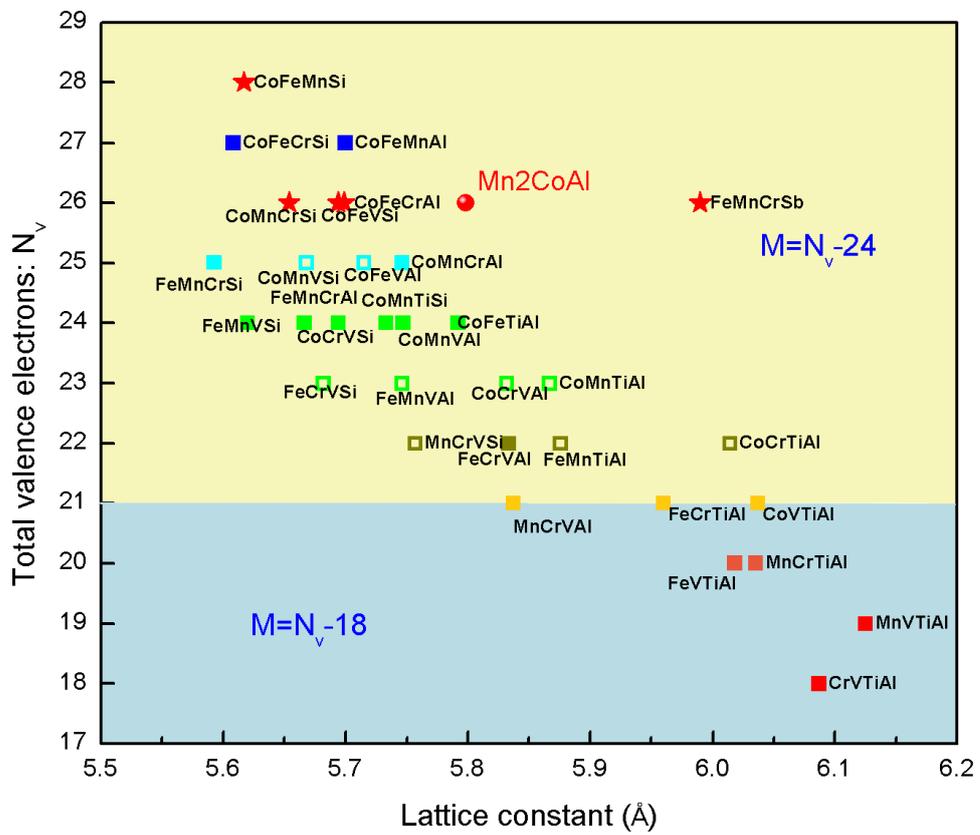

Figure 3

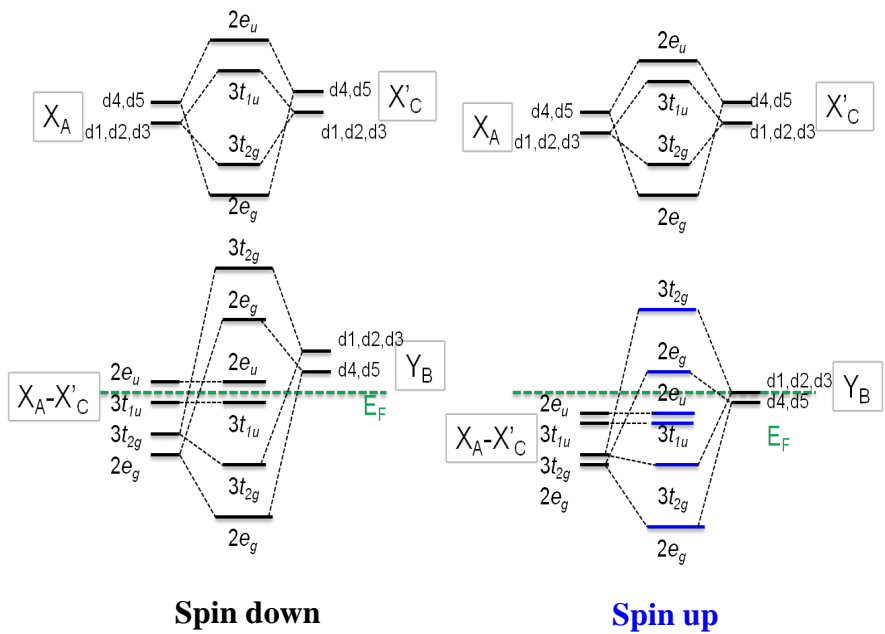

Figure 4

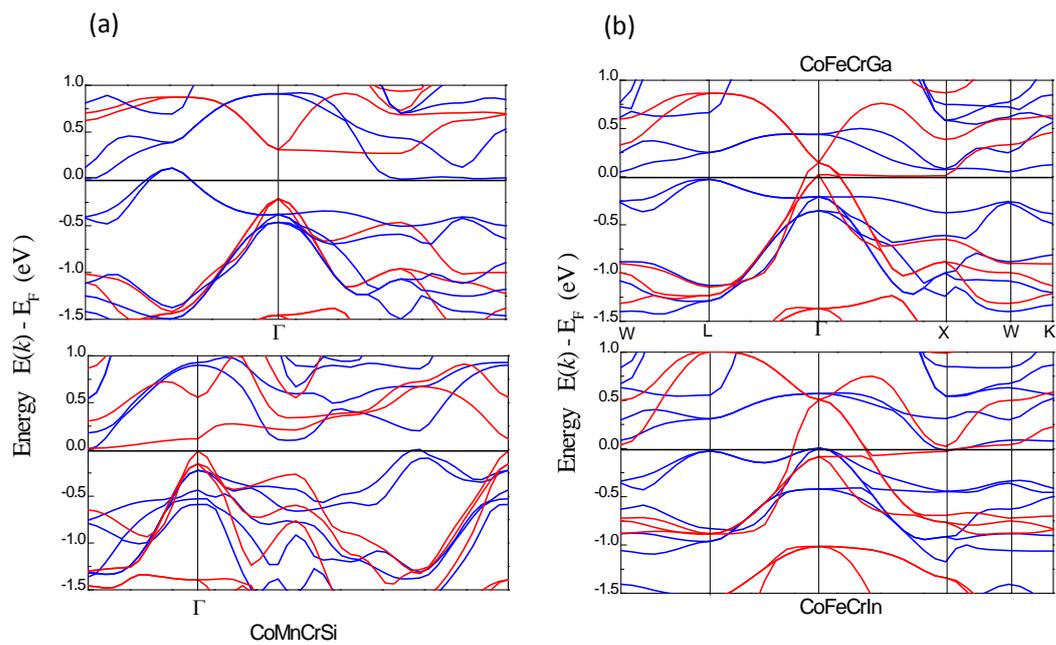

Figure 5

# A New Spin Gapless Semiconductors Family：Quaternary Heusler Compounds

G. Z. Xu[1], E. K. Liu[1], Y. Du[1], G. J. Li[1], G. D. Liu[2], W. H. Wang,[1,a)] and G. H. Wu[1]

[1]*Beijing National Laboratory for Condensed Matter Physics, Institute of Physics, Chinese Academy of Sciences, Beijing 100190, P. R. China*

[2]*School of Material Science and Engineering, Hebei University of Technology, Tianjin 300130, P. R. China*

## Supplementary Information

In table S1, we show the calculated equilibrium lattice parameters for the five compounds that are suggested to be spin gapless semiconductors. The total and atom-resolved magnetic moments of them are given in the units of $\mu_B$.

Table S1

| XX'YZ | a (Å) | $m^{total}(\mu_B)$ | $m^X$ | $m^{X'}$ | $m^Y$ | $m^Z$ |
|---|---|---|---|---|---|---|
| CoFeMnSi | 5.62 | 4 | 0.88 | 0.56 | 2.54 | 0.00 |
| CoFeCrAl | 5.70 | 2 | 0.80 | -0.64 | 1.84 | 0.00 |
| CoFeVSi | 5.69 | 1.95 | 1.06 | 0.72 | 0.18 | 0.00 |
| CoMnCrSi | 5.65 | 2 | -0.72 | 1.36 | -2.68 | 0.04 |
| FeMnCrSb | 5.99 | 2 | 0.28 | -1.56 | 3.32 | -0.02 |

Figure S1 present the spin-resolved density of states of CoFeMnSi, CoMnCrSi,CoFeVSi and FeMnCrSb . The positive DOS with blue filled area correspond to the majority spin state, the downward part with red color correspond to the minority spin state. The line locates at the energy axis of zero is the Fermi level. It can be seen that the DOS of the four compounds all reveal a gap around the Fermi energy in the minority-spin direction and a valley approaching zero at the Fermi level in the majority spin direction.

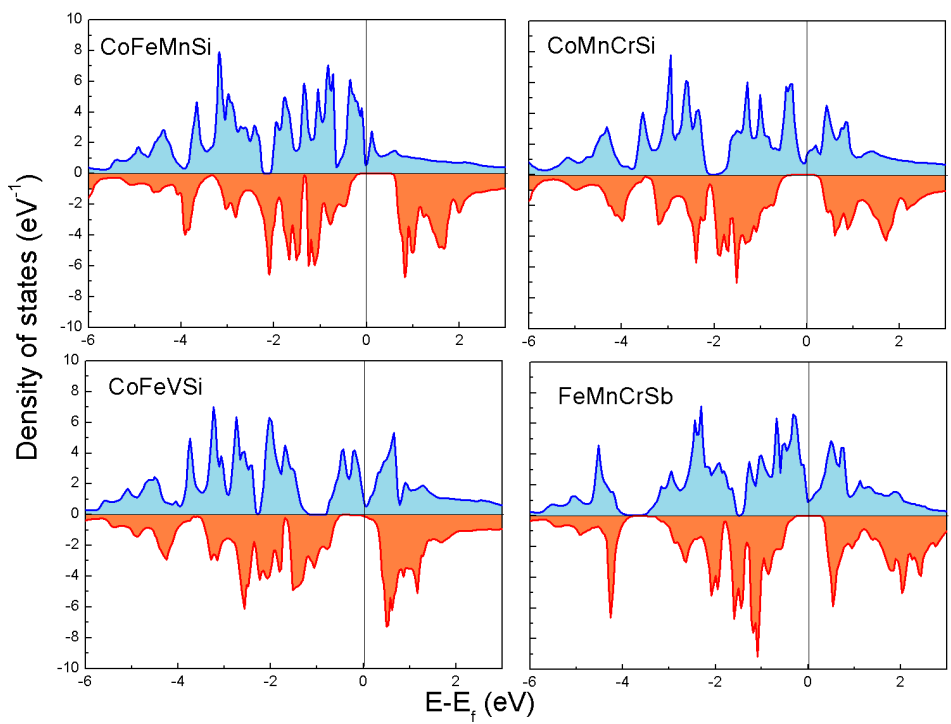

Figure S1